\documentclass[aps,pra,twocolumn,superscriptaddress,showpacs]{revtex4-1}

\usepackage[utf8]{inputenc}
\usepackage[english]{babel}
\usepackage{graphicx}
\usepackage{amsmath}
\usepackage[hidelinks]{hyperref}

\def\beq{\begin{equation}}
\def\eeq{\end{equation}}
\def\ba{\begin{align}}
\def\enda{\end{align}}
\def\bi{\begin{itemize}}
\def\ei{\end{itemize}}

\def\vPhi{\langle\Phi\rangle}

\begin{document}

\title[]{Mean-field phase diagram of ultracold atomic gases in cavity quantum electrodynamics}

\author{Lukas Himbert}
\affiliation{Theoretical Physics, Saarland University, Campus E2.6, D--66123 Saarbr\"ucken, Germany}
\author{Cecilia Cormick}
\affiliation{IFEG, CONICET and Universidad Nacional de C\'ordoba, Ciudad
Universitaria, X5016LAE C\'ordoba, Argentina}
\author{Rebecca Kraus}
\affiliation{Theoretical Physics, Saarland University, Campus E2.6, D--66123 Saarbr\"ucken, Germany}
\author{Shraddha Sharma}
\affiliation{Theoretical Physics, Saarland University, Campus E2.6, D--66123 Saarbr\"ucken, Germany}
\author{Giovanna Morigi}
\email{giovanna.morigi@physik.uni-saarland.de}
\affiliation{Theoretical Physics, Saarland University, Campus E2.6, D--66123 Saarbr\"ucken, Germany}
\date{\today}

\begin{abstract}
We investigate the mean-field phase diagram of the Bose-Hubbard model with infinite-range interactions in two dimensions.
This model describes ultracold bosonic atoms confined by a two-dimensional optical lattice and dispersively coupled to a cavity mode with the same wavelength as the lattice.
We determine the ground-state phase diagram for a grand-canonical ensemble by means of analytical and numerical methods. Our results mostly agree with the ones reported in Dogra \textit{et al.} [PRA \textbf{94}, 023632 (2016)], and have a remarkable qualitative agreement with the quantum Monte Carlo phase diagrams of Flottat \textit{et al.} [PRB \textbf{95}, 144501 (2017)]. The salient differences concern the stability of the supersolid phases, which we discuss in detail. Finally, we discuss differences and analogies between the ground state properties of strong long-range interacting bosons with the ones predicted for repulsively interacting dipolar bosons in two dimensions.
\end{abstract}

%\pacs{03.75.Hh, 37.30.+i, 32.80.Qk, 42.50.Lc}

\maketitle

\section{Introduction}

The Bose-Hubbard model is a paradigmatic quantum mechanical description of strongly-correlated spinless particles in a lattice \cite{bloch-dalibard-zwerger-2008}.
This model predicts a quantum phase transition, which emerges from the competition between the hopping coupling nearest-neighbour lattice sites and the onsite repulsion.
The latter penalizes multiple occupation at the individual sites and favours the Mott-insulator (MI) phase, while the hopping promotes the buildup of non-local correlations and the onset of superfluidity (SF) \cite{fisher-et-al-1989}.
Ultracold vapours of alkali-metal atoms offer a remarkable setup for realising and investigating the dynamics of the Bose-Hubbard model, thanks to the possibility to independently tune onsite interactions and lattice depth \cite{jaksch-et-al-1998, greiner-et-al-2002}.
The addition of further interactions, such as interparticle potentials that decay with a power-law of the inverse of the distance, typically gives rise to frustration and results in the appearance of new phases \cite{menotti-et-al-2008, li-et-al-2018, larson-et-al-2008}.
The recent realization and measurement of the phases of ultracold atomic lattices in an optical cavity has set the stage for the studies of Bose-Hubbard models with infinite-range interactions \cite{landig-et-al-2016}.
Here, attractive infinite-range interactions are realised via multiple scattering of cavity photons and favour the formation of density patterns which maximize the intracavity field \cite{domokos-ritsch-2002, asboth-et-al-2005, fernandez-vidal-et-al-2010, habibian-et-al-2013}.
When the atoms are confined by an external optical lattice and the wavelengths of the cavity mode and of the laser generating the underlying optical lattice coincide, the emerging patterns can have a checkerboard density modulation, and are denoted by charge-density wave (CDW) or lattice supersolid (SS) depending on whether the gas is incompressible or superfluid, respectively.
%The emerging ground-state phases have been extensively studied in the literature \cite{dogra-et-al-2016, niederle-morigi-rieger-2016, sundar-mueller-2016, flottat-et-al-2017, wald-et-al-2018, liao-et-al-2018, hruby-et-al-2018} and are theoretically analysed in this work.

The experimentally measured phase diagram of Refs.~\cite{landig-et-al-2016, hruby-et-al-2018} reports the existence of these four phases. %It further gives rise to a series of questions on the nature of the transitions, such as whether there is a direct transition between CDW and SF and what is the nature of the transition separating MI and CDW.
Several theoretical works reproduced the salient features of the phase diagram using different approaches.
Most works use different implementations of the mean-field treatment \cite{dogra-et-al-2016, niederle-morigi-rieger-2016, sundar-mueller-2016, wald-et-al-2018, liao-et-al-2018}, nevertheless their predictions do not agree across the whole phase diagram. Moreover, they exhibit several discrepancies with state-of-the-art two-dimensional quantum Monte Carlo (QMC) study \cite{flottat-et-al-2017}.
%Some of these discrepancies can in our view be understood from the fact that the infinite-range interactions give rise to several metastable states which hinder the search for the ground state, as we argue in this paper.
%On the other hand, infinitely-range interactions are typically well described by mean-field treatments, so that it is interesting to understand why different implementations of the mean-field treatment may lead to different results.
It has been further argued that the mean field predictions for this kind of Bose-Hubbard model shall be the same as the one of Bose-Hubbard with nearest-neighbour (and thus also repulsive dipolar) coupling \cite{dogra-et-al-2016, niederle-morigi-rieger-2016}.

The objective of this work is to provide an extensive mean-field analysis of the two-dimensional Bose-Hubbard model with infinite long-range interactions, describing the setup of Ref.~\cite{landig-et-al-2016}. For this analysis we consider a grand-canonical Hamiltonian and determine the ground-state phase diagram by taking particular care of the convergence criterion of the numerical results and by comparing them with analytical results. We then discuss our results comparing them in detail with the analytical predictions and with previous theoretical studies for the cavity Bose-Hubbard model and for repulsively interacting dipolar gases in two dimensions.

This manuscript is organized as follows. In Sec.~\ref{Sec:1} we introduce the Bose-Hubbard model and review the ground-state properties in the atomic limit, namely, when the kinetic energy is set to zero.
In Section \ref{Sec:2} we derive the mean-field Hamiltonian and employ the path-integral formalism to analytically determine the transition from incompressible to compressible phases. In Sec.~\ref{Sec:3} we numerically determine the ground-state phase diagram and compare our results with the ones reported so far in the literature.
The conclusions are drawn in Sec.~\ref{Sec:5} while the appendices provide details on the numerical methods for calculating the mean-field phase diagrams.

\section{Bose-Hubbard Hamiltonian with cavity-mediated interactions}
\label{Sec:1}

In this section we introduce the Bose-Hubbard model in presence of cavity-mediated long-range interactions and review the exact result for eigenstates and eigenspectrum in the so-called atomic limit, where the kinetic energy is set to zero.
This limit is relevant for the mean-field study, where we determine the ground state for finite values of the hopping term.

\subsection{Grand-canonical Hamiltonian}

We consider $N$ bosons tightly confined in the lowest band of a two-dimensional, quadratic optical lattice.
The lattice has $K=L\times L$ sites, each site is labeled by the index $i=(i_1,i_2)$ and we assume periodic boundary conditions.
The bosons are described by operators $\hat a_i$ and  $\hat a_i^\dagger$, which annihilate and create a particle at site $i=(i_1,i_2)$, respectively, and obey the commutation relation $[\hat a_i,\hat a_j^\dagger]=\delta_{i_1,j_1}\delta_{i_2,j_2}$.
The bosons interact via $s$-wave scattering. They also interact via the long-range interactions mediated by the photons of a single-mode cavity, whose periodicity is commensurate with the optical lattice and which gives rise to all-to-all coupling. The Hamiltonian $\hat H=\hat H_t+\hat V_0$ governing the many-body dynamics is here decomposed into the kinetic energy $H_t$ and potential energy $V_0$, which individually read \cite{habibian-et-al-2013}
\begin{eqnarray}
\label{H:tunneling}
\hat H_t &=&  - t \sum_{\langle i j \rangle} \left( \hat a_i^\dagger \hat a_j+\hat a_j^\dagger \hat a_i \right), \\
\hat V_0 &= & \frac{U_0}{2} \sum_i \hat n_i (\hat n_i - 1) -  K U_\infty  \hat \Phi^2\,,
\label{H:potential}
\end{eqnarray}
The  kinetic energy is scaled by the hopping coefficient $t$, which is positive and uniform across the lattice, and $\sum_{\langle i j \rangle}$ is restricted to the pairs of nearest neighbour sites $i$ and $j$. The potential energy is diagonal on the eigenstates of operator $\hat n_i = \hat a_i^\dagger \hat a_i$ and consists of the onsite repulsion, which is scaled by the strength $U_0>0$, and of the attractive infinite-range interactions with strength $U_\infty$, which multiplies the extensive operator $K\hat\Phi^2$. For the setup of Ref. \cite{landig-et-al-2016}, where cavity and optical-lattice laser wave lengths are equal, operator $\hat \Phi$ takes the form 
\begin{equation}
\label{eq:Phi}
\hat \Phi = \sum_j(-1)^{j} \hat n_j / K\,,
% = \sum_j (-1)^{j_1 + j_2} \hat n_j / K,
\end{equation}
where $(-1)^{j}\equiv (-1)^{j_1 + j_2}$. Its expectation value is maximum and proportional to the mean density when the atoms form a checkerboard pattern.
We denote a site $(j_1,j_2)$ by even (odd) when $j_1+j_2$ is an even (odd) number.
%The factor $K$ scaling the long-range interaction term, together with the definition \eqref{eq:Phi}, warrants the extensivity of the Hamiltonian.

In the rest of this paper we will study the phase diagram of a grand-canonical ensemble at $T=0$.
For this purpose we analyse the ground state of the grand-canonical Hamiltonian, defined as
\begin{equation}
\label{eq:GC}
\hat H_{GC}=\hat H -\mu \sum_j \hat n_j \,.
\end{equation}
Here, $\mu$ is the chemical potential which controls the mean occupation number $\rho$,
\begin{equation}
\label{eq:nbar}
\rho = \frac{1}{K} \sum_j \langle \hat n_j \rangle \,,
\end{equation}
and the expectation value is taken over the grand-canonical ensemble.
In the following we also use the parameter $\theta$, which is proportional to the expectation value of operator $\hat \Phi$ according to the relation:
\begin{equation}
\label{eq:theta}
\theta = 2 \left| \langle \hat \Phi \rangle \right|,
\end{equation}
where the proportionality factor 2 is introduced for later convenience.
The value of $\theta$ measures the population imbalance between even odd sites, thus when it is non-vanishing the atomic density is spatially modulated. In particular, it is proportional to the value of the structure form factor at the wave number of the cavity field \cite{niederle-morigi-rieger-2016}.

\subsection{Atomic limit}

We now consider the limit $t=0$. In this case the energy eigenstates are the Fock states $|n_{(1,1)},\ldots,n_{(L,L)}\rangle$, with $|n_i\rangle$ Fock state at site $i$.
It is convenient to decompose the Fock number $n_j$ of each site $j$ as the sum
\begin{equation}
\label{eq:nj}
n_j=\rho+(-1)^j\frac{\theta}{2}+\delta_j\,,
\end{equation}
where $\delta_j$ ensures that $n_j$ is a natural number.
This condition, together with Eqs. \eqref{eq:nbar} and \eqref{eq:theta}, lead to the relations
\begin{eqnarray}
  &&\sum_j \delta_j= 0\,,\\
  &&\sum_j (-1)^j \delta_j = 0\,.
\end{eqnarray}
Using these relations and Eq.\ \eqref{eq:nj} one can verify that, whenever $\rho \pm \theta/2$ is an integer number, the configuration with minimal energy has $\delta_j=0$.
In fact, using Eq.~\eqref{eq:nj} one can cast the energy of the state $|n_{(1,1)},\ldots,n_{(L,L)}\rangle$ into the form
\begin{equation}
 E( \rho, \theta, \{ \delta_j\} ) = E_0( \rho, \theta )+ \frac{U_0}{2} \sum_j \delta_j^2\,,
\end{equation}
where $ E_0( \rho, \theta )$ is the energy of the configuration when $\delta_j$ vanishes at all sites,
\begin{equation}
\label{eq:E0}
 E_0( \rho, \theta )= K \left[ \frac{U_0}{2} \rho (\rho - 1)+\left(\frac{U_0}{2} - U_\infty\right) \frac{\theta^2}{4}  - \mu \rho \right]\,,
 \end{equation}
and is visibly extensive.
This expression is correct when $|\theta| \le 2 \rho$. The ground state is found by the configuration which minimizes the energy $ E_0( \rho, \theta )$.
Therefore the ground-state properties are determined by two independent parameters, which we choose here to be $U_\infty$ and $\mu$ in units of $U_0$.
More generally,  the states at energy $E_0$ are characterised by two-site translational symmetry along both directions of the lattice, such that the sites with the same parity are equally populated.
Hence, we can denote the ground state by the ket $\{ n, m \}$ where $n$ ($m$) is the Fock number for the even (odd) sites, or vice versa.

\begin{figure}
  \includegraphics{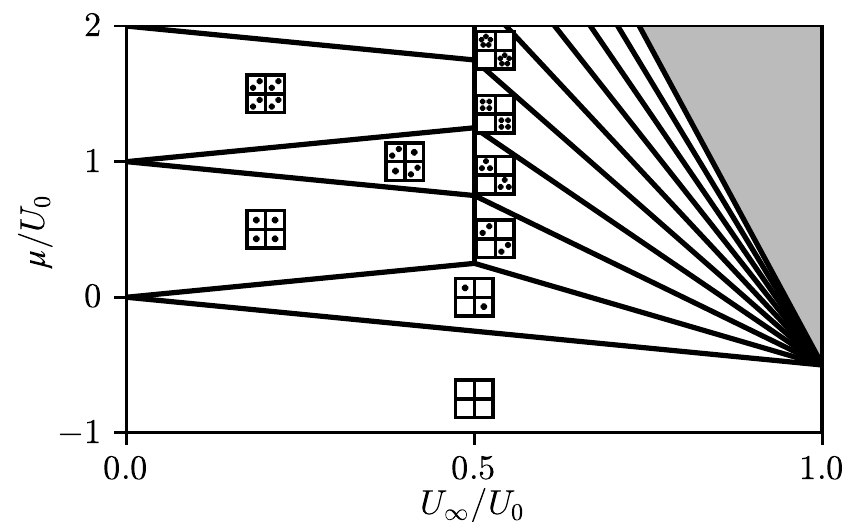}
  \caption{
  Ground-state phase diagram of the extended Bose-Hubbard model with repulsive cavity-mediated long-range interaction and repulsive on-site interaction $U_0 > 0$ in the atomic limit ($t = 0$), as a function of the chemical potential $\mu$ and the long-range interaction coefficient $U_\infty$ (both in units of $U_0$).
  The lines denote the boundaries between the incompressible phases, which are found assuming an elementary $2\times 2$ cell (indicated by the inset squares) for the CDW phase.
  The grey region contains CDW phases with increasingly large density.
  The boundaries are given by Eq.~\eqref{eq:U0/2} for $0<U_\infty<U_0/2$.
  For $U_0/2<U_\infty<U_0$ the lines correspond to Eq.~\eqref{eq:U0}.
  For $U_\infty > U_0$, the model based on the grand-canonical ensemble becomes invalid.
  }
  \label{fig:2}
\end{figure}

In the following we review the ground-state properties in the thermodynamic limit $K \to \infty$ by analysing Eq.~\eqref{eq:E0}. They can be displayed by means of a phase diagram in the $U_{\infty}-\mu$ space shown in Fig.~\ref{fig:2}, see also Ref.~\cite{dogra-et-al-2016,liao-et-al-2018,sundar-mueller-2016}.
We first notice that in the limit $U_\infty=0$ the phase is MI with commensurate density $\rho=n$ in the interval $\mu\in [U_0(n-1),U_0 n]$, while for $\mu<0$ the density is $\rho=0$.
At $\mu=U_0 n$ there is an infinite degeneracy of SF phases with density continuously varying from $n$ to $n+1$.
For increasing value of $U_\infty$, but $U_\infty < U_0/2$, the MI phase with commensurate density $n$ is the ground state for values of the chemical potential such that
\begin{equation}
\label{eq:U0/2}
U_0(n-1)+\frac{U_\infty}{2}<\mu<U_0 n-\frac{U_\infty}{2}.
\end{equation}
At the upper (lower) boundary there is an abrupt jump from the MI to a CDW phase with fractional density $n+1/2$ ($n-1/2$) and population imbalance $|\theta|=1$.
In this CDW the occupation of two adjacent sites is $\{n, n+1\}$ ($\{n, n-1\}$), or vice versa, the CDW ground state being doubly degenerate.
These boundaries are the lines depicted in Fig.~\ref{fig:2}.
At $U_\infty=U_0/2$ there is a discontinuity: For $U_\infty>U_0/2$ the ground state at density $\rho$ is a CDW with the largest population imbalance $|\theta|=2\rho$ (where $2\rho$ is an integer) in the interval
\begin{equation}
\label{eq:U0}
(U_0-U_\infty)(\theta-1)-\frac{U_\infty}{2}<\mu<(U_0-U_\infty)\theta-\frac{U_\infty}{2}\,,
\end{equation}
while at $\mu=2(U_0-U_\infty)\rho-\frac{U_\infty}{2}$ there is an infinite manifold of SF states with density varying from $\rho$ to $\rho+1/2$.
The corresponding phases and boundaries are shown in Fig.~\ref{fig:2}, they all converge to the same point at $U_\infty=U_0$.
For $U_\infty\ge U_0$ the onsite energy is attractive, the energy is not bound from below and the grand-canonical ensemble becomes unstable.

This analysis reproduces the results of Ref.~\cite{dogra-et-al-2016,liao-et-al-2018, sundar-mueller-2016} for $t=0$.
In what follows we will study the phase diagram for finite tunneling rates by means of the mean-field analysis.

\section{Mean-field analysis}
\label{Sec:2}

In this Section we review the mean-field model which is at the basis of our numerical calculations and the definition of the observables that identify the phases.
We then use the path-integral formalism to analytically determine the transition from incompressible to SF phase.

\subsection{Mean-field treatment}

We first introduce a so-called "local superfluid order parameter", which is the expectation value of the annihilation operator $\hat{a}_i$ at site $i$:
\begin{equation}
  \varphi_i = \langle \hat{a}_i \rangle.
\end{equation}
The mean-field approximation consists of neglecting terms in second-order in the fluctuations $\delta \hat{a}_i$ of the annihilation operator about $\varphi_i$, with $\delta \hat{a}_i = \hat{a}_i - \varphi_i$.
With this approximation the Hamiltonian term \eqref{H:tunneling} can be cast into the sum of local Hamiltonians $\hat H_t^{\text{mf}}=\sum_i \hat H_t^{(i)}$ with
\begin{equation}
\hat H_t^{(i)}= - t \left( \hat a_i^\dagger \bar{\varphi}_i + \hat a_i \bar{\varphi}_i^* - {\rm Re}\{\varphi_i^* \bar{\varphi}_i\}\right)\,,
\end{equation}
and where $\bar{\varphi}_i = \sum_{\langle j \rangle_i} \varphi_j$ is the sum of local SF order parameters of the neighbours of site $i$. Without loss of generality in the
numerical calculations we assume that these parameters are real.% (any complex value can be reduced to a real one by defining a unitary transformation that rotates the operators $a_i$).

In order to write the total Hamiltonian in terms of local operators, we perform a second approximation by writing the cavity potential in the mean-field form: $\hat\Phi^2\approx \theta\hat\Phi-\theta^2/4$, thus we discard fluctuations of $\hat\Phi$ to second order. With this approximation we can now write the grand-canonical Hamiltonian, Eq.~\eqref{eq:GC}, in its mean field form $H_{\rm GC:mf}=\sum_i\hat H^{(i)}_{\rm mf}$, namely, as the sum of local-site Hamiltonians $\hat H^{(i)}_{\rm mf}$ that read
\begin{equation}
\hat H^{(i)}_{\rm mf}=\hat H_t^{(i)} + \frac{U_0}{2} n_i (n_i - 1)
   - (-1)^i U_\infty \theta n_i + U_\infty \frac{\theta^2}{4} - \mu n_i\,.
\end{equation}
In the following we assume two-site symmetry, as in Ref.~\cite{dogra-et-al-2016}.
%This is a reliable assumption for the ground state, as we argued in the previous section.
Using this assumption all even and all odd sites possess the energy $\hat H^{(e)}_{\rm mf}$ and $\hat H^{(o)}_{\rm mf}$, respectively, such that $\hat H_{\rm GC:mf}=K(\hat H^{(e)}_{\rm mf}+\hat H^{(o)}_{\rm mf})/2$.
It is now convenient to introduce the annihilation and creation operators $\hat a_e$ and $\hat a_e^\dagger$ ($\hat a_o$ and $\hat a_o^\dagger$) for a particle in an even (odd) site, and the corresponding number operator $\hat n_e=\hat a_e^\dagger\hat a_e$ ($\hat n_o=\hat a_o^\dagger\hat a_o$).
The even (odd) sites have SF order parameter $\varphi_e$ ($\varphi_o$) and the population imbalance operator reads $\hat \Phi=(\hat n_e-\hat n_o)/2$.
With these definitions we write
\begin{equation}
  \begin{split}
  H_{s \in \{ \text{e}, \text{o} \}} = & - z t \varphi_{\bar s} (\hat{a}_s + \hat{a}_s^\dagger - \varphi_{s}) + \frac{U_0}{2} \hat{n}_s (\hat{n}_s - 1) \\
  & - \mu \hat{n}_s- \sigma_s U_\infty \theta \hat{n}_s +U_\infty \theta^2/4\ \text{,}
  \end{split}
  \label{eq:mean-field-hamiltonians}
\end{equation}
where we have used that $\bar\varphi_s=z\varphi_{\bar s}$, with $z$ the coordination number (here equal to 4) and  $\varphi_{\bar e} = \varphi_o$ ($\varphi_{\bar o} = \varphi_e$). Moreover, we have introduced the symbol $\sigma_{\text{e}} = +1$, $\sigma_{\text{o}} = -1$. Hamiltonian \eqref{eq:mean-field-hamiltonians} is at the basis of the numerical results presented in the next section.

\subsection{Transition from incompressible to compressible phases}

We now determine the critical tunneling rate which separates compressible from incompressible phases.
For this purpose we start from Hamiltonian \eqref{eq:GC} and consider an elaborate form of mean-field treatment following Refs.~\cite{kampf-zimanyi-1993, fisher-et-al-1989, sachdev-2011}.
We consider the partition function \cite{muehlschlegel-1978, fisher-et-al-1989}:
\beq
Z = {\rm Tr} \left\{ e^{-\beta \hat H_0} T_\tau e^{-\int_0^\beta d\tau \hat H_I(\tau)} \right\}
\label{eq:Z1}
\eeq
where $\beta$ is the inverse temperature, $\tau$ is the imaginary time, $T_\tau$ is the imaginary-time ordering operator, $\hat H_0=\hat V_0-\mu\sum_i\hat n_i$ is the grand-canonical Hamiltonian without the kinetic energy, and we take $\hbar =1$ to simplify the notation.
Moreover,
\beq
\hat H_I(\tau) = e^{\tau \hat H_0} \hat H_t e^{-\tau \hat H_0}\,,
\label{eq:imaginary Heisenberg}
\eeq
where $\hat H_t$ is the tunneling Hamiltonian, Eq.~\eqref{H:tunneling}.
We can also write Equation (\ref{eq:Z1}) as
%Equation (\ref{eq:Z1}) can also be written as 
$Z = Z_0 \left\langle T_\tau e^{-\int_0^\beta d\tau \hat H_1(\tau)} \right\rangle_0$
 \cite{fisher-et-al-1989}, where $Z_0$ the partition function for the model corresponding to $\hat H_0$ and the expectation value evaluated for the thermal state of the same model at inverse temperature $\beta$.
Equivalently, one can cast the expression in terms of coherent-state path integrals \cite{kampf-zimanyi-1993, sachdev-2011, negele-orland-1998}:
\beq
\label{eq:Z2}
Z = \int \mathcal D \alpha_j \mathcal D \alpha_j^* e^{-\int_0^\beta d\tau \mathcal L (\tau)}
\eeq
where
\beq
\mathcal L = \sum_j \alpha_j^* \frac{d\alpha_j}{d\tau} +H(\{\alpha_j^*, \alpha_j\} ) \,.
\eeq
where $H$ is assumed to be written in normal form and the path integral is over variables satisfying periodic boundary conditions.
The two formalisms can be related by noting that for an operator $A[\{\hat a_j^\dagger (\tau_j), \hat a_{j'} (\tau_{j'})\} ]$ \cite{negele-orland-1998}:
\begin{multline}
\langle T_\tau A[\{\hat a_j^\dagger (\tau_j), \hat a_{j'} (\tau_{j'})\} ] \rangle_0 = \\
\frac{1}{Z_0} \int \mathcal D \alpha_j \mathcal D \alpha_j^* e^{-\int_0^\beta d\tau \mathcal L (\tau)} A[ \{\alpha_{j'}^* (\tau_{j'}), \alpha_j(\tau_j)\} ]\,,
\end{multline}
where the imaginary time dependence of the operators is defined in the same way as in Eq.~(\ref{eq:imaginary Heisenberg}).

We define a new basis of Fourier-transformed variables $\alpha_q, \alpha_q^*$, with $q=(q_1,q_2)$:
\beq
\alpha_q = \frac{1}{\sqrt{K}} \sum_j \alpha_j \exp[2\pi i (j_1 q_1+j_2 q_2)/\sqrt{K}]
\label{eq:Fourier}
\eeq
with  $L=\sqrt{K}$. We then write
\beq
\mathcal L = \mathcal L_0 - t \sum_q w_q \alpha_q^* \alpha_q  \ ,
\eeq
where $w_q=2(\cos{(2\pi q_1/\sqrt{K})}+\cos(2\pi q_2/\sqrt{K}))$ are the eigenvalues of the vicinity matrix, and where $\mathcal L_0$ is the Lagrangian without the tunneling terms.
By means of the Hubbard-Stratonovich transformation we can write
\beq
e^{t \int d\tau \sum_q w_q \alpha_q^* \alpha_q} = \int \mathcal D \psi_q \mathcal D \psi_q^* e^{- \int d\tau \mathcal L_2 + \mathcal L_c} \ ,
\eeq
where all normalization factors are now included in the definition of the functional integral, and
\begin{eqnarray}
\mathcal L_2 &=& t \sum_q \psi_q^* \psi_q \,,\\
\mathcal L_c &=& -t \sum_q \sqrt{w_q} (\alpha_q^* \psi_q + \psi_q^* \alpha_q)\,.
\end{eqnarray}
The prefactors here are chosen so that $\psi_q$ are dimensionless.
In particular, the auxiliary variables $\psi_q, \psi_q^*$ are related to the Fourier transform of the expectation values $\varphi_i$ by the equation
$\langle \psi_q \rangle = \sqrt{w_q} \varphi_q$.

We now integrate Eq.~\eqref{eq:Z2} over the variables $\alpha_j, \alpha_j^*$ and obtain
\beq
Z = Z_0 \int \mathcal D \psi_q \mathcal D \psi_q^* e^{-S_{\rm eff}}\,,
\eeq
where we have introduced the effective action $S_{\rm eff}$.
The effective action is non-local in time and is given by the expression
\begin{eqnarray}
S_{\rm eff} &=& - \ln \Big( \frac{1}{Z_0} \int \mathcal D \alpha_j \mathcal D
\alpha_j^* e^{-\int_0^\beta d\tau \mathcal L_0 + \mathcal L_c} \Big) +
\int_0^\beta d\tau \mathcal L_2 \quad\quad
\nonumber\\
&=& - \ln\left( \langle e^{-\int_0^\beta d\tau \mathcal L_c} \rangle_0 \right) +
\int_0^\beta d\tau \mathcal L_2\,.
\end{eqnarray}
In order to find the transition points, it is sufficient to consider $S_{\rm eff}$ up to second order in the auxiliary fields. One recovers the expression \cite{kampf-zimanyi-1993}:
\beq
\label{Seff:2}
S_{\rm eff}^{(2)} = - \frac{1}{2} \Big\langle \Big(\int_0^\beta d\tau \mathcal L_c\Big)^2 \Big\rangle_0 + \int_0^\beta d\tau \mathcal L_2
\eeq
Owing to the phase invariance of the model, Eq. \eqref{Seff:2} reduces to the form:
\begin{multline}
S_{\rm eff}^{(2)} = \int_0^\beta \!\!\! d\tau \mathcal L_2 - t^2 \sum_{q, q'} \sqrt{w_q w_{q'}} \int_0^\beta \!\!\! d\tau \int_0^\tau \!\!\! d\tau' \\\Big[ \psi_q^*(\tau) \psi_{q'} (\tau') \langle  \hat a_q(\tau)  \hat a_{q'}^\dagger (\tau') \rangle_0 +  \psi_q(\tau) \psi_{q'}^* (\tau') \langle  \hat a^\dagger_q(\tau)  \hat a_{q'} (\tau') \rangle_0 \Big] \,,
\label{eq:Seff}
\end{multline}
where the Fourier-transformed operators $\hat a_q$ of the site operators $\hat a_j$ are defined in analogous form as Eq.~\eqref{eq:Fourier}.

The time correlators of the model with no hopping can be calculated easily in the site basis.
For the case $T\to0$ ({\it i.e.} $\beta\to\infty$) they are found to be:
\begin{eqnarray}
\label{Corr:1}
\langle \hat a_j^\dagger(\tau) \hat a_{j'} (\tau-\tau_0) \rangle_0 &=& \delta_{jj'} n_j e^{-t E_j^-} \,, \\
\langle \hat a_j(\tau) \hat a_{j'}^\dagger (\tau-\tau_0) \rangle_0 &=& \delta_{jj'} (n_j+1) e^{-t E_j^+}\,,
\label{Corr:2}
\end{eqnarray}
where the values of $n_j$ and $\vPhi$ are the ones that correspond to the ground state for $t=0$, see Sec.~\ref{Sec:1}. The energy $E_j^\pm$ is the energy variation resulting from the addition or subtraction of a particle at site $j$,
\begin{eqnarray}
E_j^- &=& \mu-U_0(n_j-1)+2U_\infty\vPhi(-1)^j \,,\\
E_j^+ &=& -\mu+U_0n_j-2U_\infty\vPhi(-1)^j
\end{eqnarray}
where we neglected a term of order $1/K$ (note that $E_j^-$ is defined for $n_j >0$).

In the ground state, $n_j$ and $E_j^\pm$ only depend on the parity of the site, so one can cast the correlators of Eqs. \eqref{Corr:1} and \eqref{Corr:2} in the form:
\begin{eqnarray}
\langle \hat a_j^\dagger(\tau) \hat a_j (\tau-\tau_0) \rangle_0 &=& C_{s\in \{e,o\}}^-(\tau_0) \,, \\
\langle \hat a_j(\tau) \hat a_j^\dagger (\tau-\tau_0) \rangle_0 &=& C_{s\in \{e,o\}}^+(\tau_0)
\end{eqnarray}
where the subindices correspond to $j$ being even or odd.
This can be used to calculate the Fourier-transformed correlators:
\begin{eqnarray}
\langle \hat a_q^\dagger(\tau) \hat a_{q'} (\tau-\tau_0) \rangle_0 &=& C_e^-(\tau_0) \frac{\delta_{qq'}+\delta_{q\bar q'}}{2} +C_o^-(\tau_0) \frac{\delta_{qq'} - \delta_{q\bar q'}}{2} \,, \quad\quad \\
\langle \hat a_q(\tau) \hat a_{q'}^\dagger (\tau-\tau_0) \rangle_0 &=& C_e^+(\tau_0) \frac{\delta_{qq'}+\delta_{q\bar q'}}{2} + C_o^+(\tau_0) \frac{\delta_{qq'} - \delta_{q\bar q'}}{2} \,. \quad\quad
\end{eqnarray}
Here, we introduced the notation:
\beq
\bar q = (\bar q_1, \bar q_2) = (q_1 + \sqrt{K}/2, q_2 + \sqrt{K}/2)
\label{eq:barq}
\eeq
and the sum of quasimomenta is taken to be mod $\sqrt{K}$.
Thus, the presence of an even-odd asymmetry leads to non-vanishing correlators between momenta corresponding to $q, \bar q$.
Note that for temperatures $T>0$ this structure is maintained, only the form of the single-site correlators is changed.
\begin{widetext}
The correlators in Fourier basis can then be replaced in expression (\ref{eq:Seff}), and the sum can be made more compact by noting that $w_{\bar q} = - w_q$.
We now make the standard assumption that the transition can be found by considering time-independent auxiliary fields $\psi_q, \psi_q^*$, so that they can be taken out of the integrals. We obtain:
\begin{eqnarray}
S_{\rm eff}^{(2)} &=&  \sum_q \psi_q^* \psi_q \Bigg\{ t \beta - \frac{t^2 w_q}{2} \int_0^\beta \!\!\! d\tau \int_0^\tau  \!\!\! d\tau_0 \, [C_{e}^-(\tau_0) + C_{o}^-(\tau_0) + C_{e}^+(\tau_0) + C_{o}^+(\tau_0) ] \Bigg\}\nonumber\\
&+& i \sum_q \psi_{\bar q}^* \psi_q \frac{t^2 w_q}{2} \int_0^\beta \!\!\! d\tau \int_0^\tau  \!\!\! d\tau_0 \, [C_{e}^-(\tau_0) - C_{o}^-(\tau_0) + C_{e}^+(\tau_0) - C_{o}^+(\tau_0) ] \,.
\end{eqnarray}

For the case $T\simeq0$, after performing the time integrals one gets:
\begin{eqnarray}
\!\!\!\! S_{\rm eff}^{(2)} &\simeq& t\beta \sum_q \Bigg\{ \psi_q^* \psi_q \Big[ 1 - \frac{t w_q}{2} \Big(\frac{n_{e}}{E_e^-} + \frac{n_{o}}{E_o^-} + \frac{n_{e}+1}{E_e^+} + \frac{n_{o}+1}{E_o^+} \Big) \Big] \nonumber\\
&  &+ i \psi_{\bar q}^* \psi_q \frac{t w_q}{2} \Big(\frac{n_{e}}{E_e^-} - \frac{n_{o}}{E_o^-} + \frac{n_{e}+1}{E_e^+} - \frac{n_{o}+1}{E_o^+} \Big) \Bigg\} \,.
\label{eq:Seff2hopping}
\end{eqnarray}
\end{widetext}
Thus, for each pair of modes $q, \bar q$, the effective action to second order has eigenvalues corresponding to a matrix of the form:
\beq
M_q = \mathbf{I} - \frac{t w_q}{2} (\ell_1 \sigma_z + i \ell_2 \sigma_x) \,,
\eeq
with $\ell_1, \ell_2$ the ($q$-independent) coefficients in Eq.~\eqref{eq:Seff2hopping}.
The smallest of each pair of eigenvalues reads $1 - t|w_q|\sqrt{\ell_1^2 - \ell_2^2}/2$. Hence, since the  largest value of $|w_q|$ is equal to 4, the smallest eigenvalue of all pairs of modes in 2 dimensions is then $1 - 2t \sqrt{\ell_1^2 - \ell_2^2}$. The transition point is found when this eigenvalue vanishes. After replacing the coefficients $\ell_1, \ell_2$ one finds:
\beq
t_c^{-1} = 4 \sqrt{\left( \frac{n_{e}}{E_e^-} + \frac{n_{e}+1}{E_e^+} \right) \left( \frac{n_{o}}{E_o^-} + \frac{n_{o}+1}{E_o^+} \right)} \,.
\label{eq:critical-tunneling}
\eeq
This result coincides with the one reported in Refs.~\cite{dogra-et-al-2016, sundar-mueller-2016,liao-et-al-2018}.

We conclude this section by remarking that this formalism should also allow one to identify the transition between SF and lattice SS. We have applied it in fact by treating the cavity potential as perturbation of the SF ground state and further performed a higher-order expansion of the effective action in order to analyse the effect of coupling between order parameters. The phase boundary we obtain, however, does not agree with the numerical results of the following section. For this reason we refrain to report the details of the calculation.

\section{Ground-state phase diagram}
\label{Sec:3}

In this section we determine the ground-state phase diagram using the mean-field model, Eq.~\eqref{eq:mean-field-hamiltonians}.
By rescaling the energy with $U_0$, the ground state is fully characterized by three parameters: $\mu$, which controls the density, $U_\infty$, which scales the cavity interactions, and the tunneling $t$.
The numerical analysis is performed by identifying self-consistently the ground state using a fixed-point iteration detailed in Appendix \ref{app:numerical-ground-state}.
By these means we identify four possible phases: (i) SF when $\varphi_s\neq 0$ and $\theta=0$; (ii) lattice supersolid (SS) when $\varphi_s\neq 0$ and $\theta\neq 0$; (iii) MI when $\varphi_s= 0$ and $\theta=0$; and finally (iv) CDW, when $\varphi_s= 0$ and $\theta\neq 0$ \cite{niederle-morigi-rieger-2016}.
We further note that in the SS phase the two SF order parameters $\varphi_{e}$ and $\varphi_{o}$ take different non-vanishing values.

\subsection{Ground-state phase diagram for varying density}

\begin{figure}
  \includegraphics{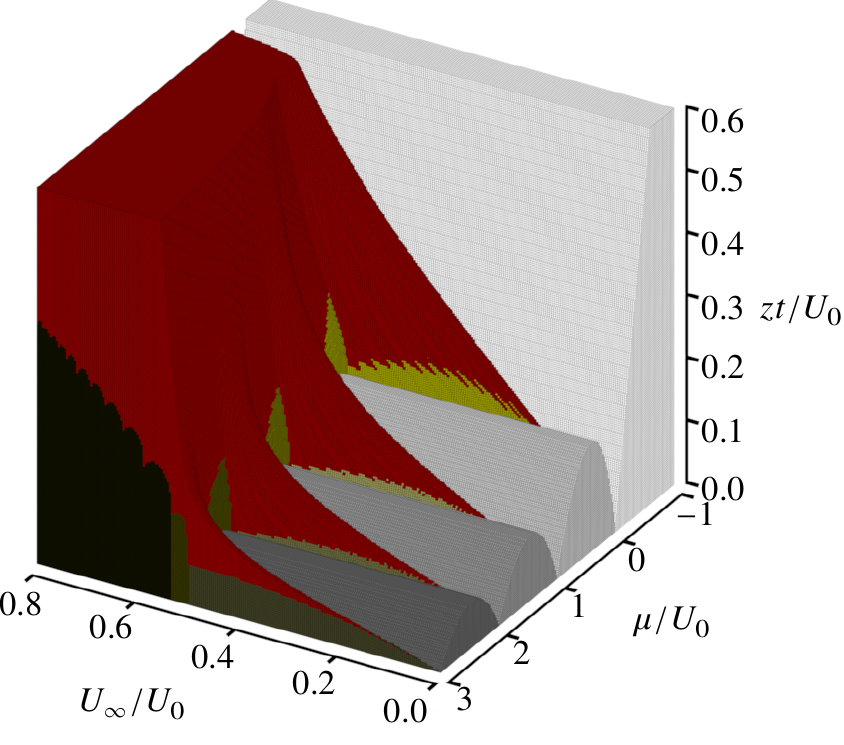}
  % File cavity-meanfield-groundstate-independent-hamiltonian-parameters-2017-12-08T10:10:41+0100.json.xz
  \caption{
  Ground-state phase diagram as a function of $U_\infty,\mu,t$ (in units of $U_0$) obtained by numerically determining the phases using the mean-field model of Eq.~\eqref{eq:mean-field-hamiltonians}.
  The MI phase is grey, the CDW is yellow, the SS is red, the rest of the phase diagram is SF.  Note that the tunneling rate is rescaled by the coordination number $z$ (here $z=4$).
  In the numerical procedure the cut-off at the occupation at each site is at $n_{\text{max}} = 31$, the precision is $\varepsilon = 10^{-8}$ and $275$ initial guesses were taken (see Appendix \ref{app:numerical-ground-state}).
  }
  \label{fig:cavity-ground-state-phase-diagram}
\end{figure}
Figure~\ref{fig:cavity-ground-state-phase-diagram} shows the ground-state phase
%Figure~\ref{fig:cavity-ground-state-phase-diagram} displays the phase
 diagram as a function of $U_\infty,\mu,t$, the different colors identify a different phase, the SF phase is the corresponding empty space.
In the plane at $U_\infty=0$ we recover the mean-field phase diagram of the two-dimensional Bose-Hubbard model \cite{fisher-et-al-1989, zwerger-2003}.
For $0 < U_\infty / U_0 < 0.5$ the MI lobes shrink along the $\mu$ axis and are sandwiched by CDW phases, which become increasingly visible.
Here, the CDW phases are characterized by minimal population imbalance $\theta=1$, corresponding to $n_e=n$ and $n_o=n+1$ where $n$ is integer, or vice versa.
The red regions at the tip of each CDW lobe is SS, the parameter region where the SS phase is different from zero increases with $U_\infty$. 

Inspecting Fig.~\ref{fig:cavity-ground-state-phase-diagram} we observe
%By inspecting Fig.~\ref{fig:cavity-ground-state-phase-diagram} we observe
 that the MI phases vanish at $U_\infty=0.5U_0$ also for finite tunneling.
Moreover, there is a discontinuity at $U_\infty=0.5U_0$: the CDW phases with population $\{ n, n+1 \}$ completely disappear and are replaced by CDW phases with maximal population imbalance $\{ 0, 2n +1\}$.
This result is in agreement with our analysis in the atomic limit, Sec.~\ref{Sec:1}.
Moreover, at $U_\infty=0.5U_0$ and at finite tunneling rate one observes a discontinuous transition from SF to CDW.
For $U_\infty>0.5U_0$ the CDW phases are separated from the SF phase by lattice SS phases, which almost completely surround the tip of CDW regions. %{\bf These observations inferred from Fig.~\ref{fig:cavity-ground-state-phase-diagram} is in good aggreement with \cite{dogra-et-al-2016, sundar-mueller-2016}}

\begin{figure}
  \includegraphics{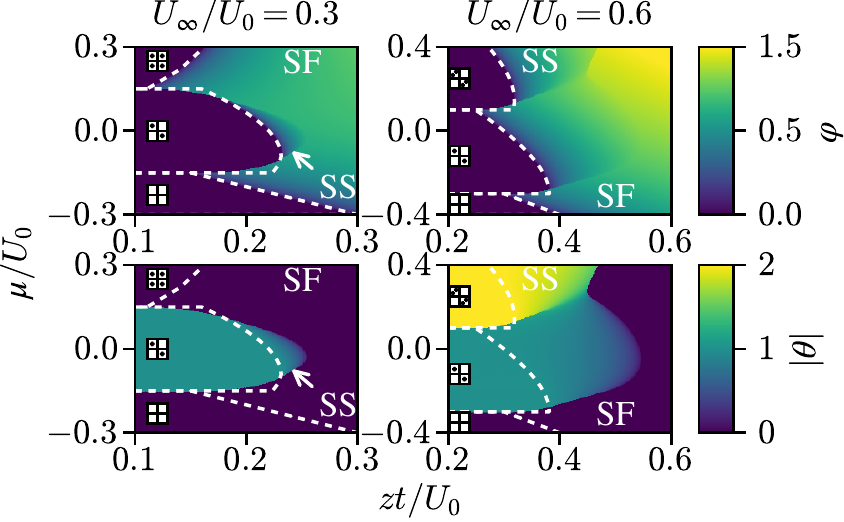}
  % See file cavity-meanfield-groundstate-independent-hamiltonian-parameters-2018-09-21T18:58:35+0200.json.xz
  \caption{
  Cuts of the ground-state phase diagram of Fig.~\ref{fig:cavity-ground-state-phase-diagram} for $U_\infty=0.3U_0$ (left column) and $U_\infty=0.6U_0$ (right column).
  Upper panels: contour plots of the SF order parameter $\varphi = \sqrt{|\varphi_{\text{e}} \varphi_{\text{o}}|}$; Lower panels: contour plots of the population imbalance $|\theta|$.
  The white dashed lines show the phase boundaries predicted by Eq. \eqref{eq:critical-tunneling}.
  }
  \label{fig:cavity-ground-state-supersolid-closeup}
\end{figure}

We now consider the values $U_\infty=0.3U_0$ and $U_\infty=0.6U_0$ and show the behavior of SF order parameter and $\theta$, respectively, in Fig.~\ref{fig:cavity-ground-state-supersolid-closeup}. %The values of $U_\infty$ chosen intentionally below and above the threshold value $U_\infty=U_0/2$. } 
%Figures~\ref{fig:cavity-ground-state-supersolid-closeup} show the SF order %parameter and $\theta$, respectively, as a function of $\mu$ and $t$ for
% $U_\infty=0.3U_0$ and $U_\infty=0.6U_0$, namely, below and above the
% threshold value $U_\infty=U_0/2$. 
We first discuss the case $U_\infty=0.3U_0$, namely, when the strength of the long-range interaction is below the threshold value $U_\infty=U_0/2$. In this case, the MI phases are stable. The transition between MI-SF and CDW-SS are characterized by a continuous change of the SF order parameter. However, there is no direct transition between the MI and SS phases. Our numerical results, moreover, predict a direct transition between CDW and MI at $t>0$. The  population imbalance changes discontinuously across the CDW-MI transition boundary.
%Here, the  population imbalance changes discontinuously. 
In particular, in the vicinity of the transitions between each two insulating lobes, we find a range of parameters where they are metastable: The transition line here corresponds to the parameters where the two states have the same energy. This prediction agrees with the ones of Refs.~\cite{dogra-et-al-2016, sundar-mueller-2016, panas-kauch-byczuk-2017}. A direct CDW-MI transition is also predicted by a mean-field treatment in a canonical ensemble \cite{wald-et-al-2018}. 

We remark that a direct CDW-MI transition, a direct CDW-SF transition, and SS phases at the tip of the CDW lobes have also been found
in mean-field studies based on cluster analysis \cite{niederle-morigi-rieger-2016}. A further quantitative
comparison with the phase diagram reported there is not possible. In fact, the effective strength
of the long-range interaction term is not constant across the phase diagram of Ref. \cite{niederle-morigi-rieger-2016}, since
this depends on the overlap integral between the cavity standing wave and the Wannier functions.
There, the Wannier functions are calculated by changing the depth of the confining optical lattice, after which the
integrals giving $t$, $U_0$, and $U_\infty$ are determined.
%This is not verified in the mean-field phase diagram of dipolar bosons \cite{menotti-trefzger-lewenstein-2007}, where CDW and MI are separated by a SF phase.

We now discuss the phase diagrams in Fig.~\ref{fig:cavity-ground-state-supersolid-closeup} for $U_\infty=0.6U_0$. Comparison with the left panels show that now the CDW lobes have moved towards smaller chemical potentials, their width (with respect to the chemical potential) has decreased and the critical tunneling rate has increased. The form of this phase diagram qualitatively agrees with the one reported in \cite{dogra-et-al-2016, sundar-mueller-2016}. Also in this case, discrepancies between the numerics and the analytical lines are visible at the direct transition CDW-SF and at the transition between CDW and SS with different values of the population imbalance. At these phase boundaries the population imbalance varies discontinuously. We discuss in the next section the nature of the other transitions. 

\begin{figure}
  \includegraphics{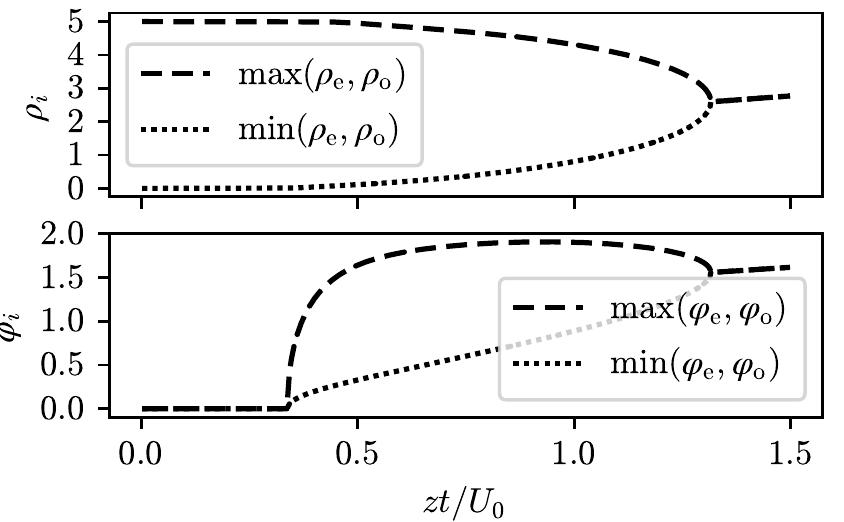}
  % See file cavity-meanfield-groundstate-independent-hamiltonian-parameters-2018-01-29T13:45:30+0100.json.xz
  \caption{
  Local density $\rho_{i}$ and SF order parameter $\varphi_{i}$ as function of the tunneling $z t / U_0$ with $z = 4$.
  The order parameters are obtained by self-consistently diagonalizing the Hamiltonians \eqref{eq:mean-field-hamiltonians} according to the procedure detailed in Appendix \ref{app:numerical-ground-state}.
  The parameters are $\mu =U_0$ and $U_\infty= 0.7U_0$.
  The occupation at each site is cut off above $n_{\text{max}} = 31$, the precision is $\varepsilon = 10^{-7}$ and $275$ initial guesses were taken.
  }
  \label{fig:cavity-no-structural-phase-transition}
\end{figure}

Finally, we observe that at fixed chemical potential and at fixed values of $U_\infty$ above $U_0/2$ the CDW phase has constant population imbalance.
This contrasts with the prediction of Ref.~\cite{liao-et-al-2018}, where a transition between CDW phases with maximal population imbalance as a function of the tunneling rate was reported.
Figure~\ref{fig:cavity-no-structural-phase-transition} shows the occupations $\rho_e$ and $\rho_o$ of the even and odd sites, respectively, as well as the corresponding SF order parameters as a function of the tunneling rate for the same parameters as in Fig.~6 of Ref.~\cite{liao-et-al-2018}.
We find that in the incompressible phase the population imbalance is constant and equal to $|\theta| = 5$.
We note that our self-consistent analysis at $t=0$ gives that the CDW with occupations $\{ 0, 4 \}$ is metastable  with energy $-1.8 U_0$, while the CDW with occupations $\{ 0, 5 \}$ is the ground state with energy $-1.85 U_0$.
Since the mean-field energy does not depend on the tunneling rate in the incompressible phase, then the $\{ 0, 5 \}$ CDW is the ground state for all values of $t$ where it is stable.
We conclude that a transition like that reported in reference \cite{liao-et-al-2018} is not consistent within the static mean field assumption.

Before concluding this section, we briefly compare the phase diagrams in Fig. \ref{fig:cavity-ground-state-supersolid-closeup} for $U_\infty=0.3U_0$ with the ones for dipolar gases, interacting repulsively in two dimensions. Here, mean-field treatments and quantum Monte Carlo calculations report the same phases as for all-to-all coupling, however the ground state phase diagrams are qualitatively different. An important difference is that for dipolar gases there is no direct transition CDW-MI \cite{kovrizhin-et-al-2005, menotti-trefzger-lewenstein-2007, iskin-2011, ohgoe-et-al-2012, batrouni-scalettar-2000}.

\subsection{Ground state for fixed densities}
\label{Sec:Density}

We now discuss the phase diagram as a function of $t/U_0$ and $U_\infty/U_0$ at fixed density $\rho$. Within our grand-canonical model this implies to find the values of the chemical potential $\mu$, at given  $\tilde t=t/U_0$ and $\tilde U_\infty=U_\infty/U_0$, which satisfy the equation $\rho(\mu/U_0,\tilde t,\tilde U_\infty)=$ constant.
%This is done using the grand-canonical ensemble by taking the values at the chemical potential $\mu$ such that the density $\rho$ is constant.
Since the compressibility shall fulfil $\partial \rho / \partial \mu \geq 0$, we can use a bisection algorithm to efficiently find the chemical potential which corresponds to a fixed density. The details are reported in Appendix \ref{app:constant-density}.
This procedure did not provide a solution for all values of parameters $t / U_0$ and $U_\infty / U_0$ because the compressibility $\partial \rho / \partial \mu$ is not continuous over the full range of $\mu$ values:  we find jumps in the density as a function of the chemical potential, as we discuss in what follows.

\begin{figure}
  \includegraphics{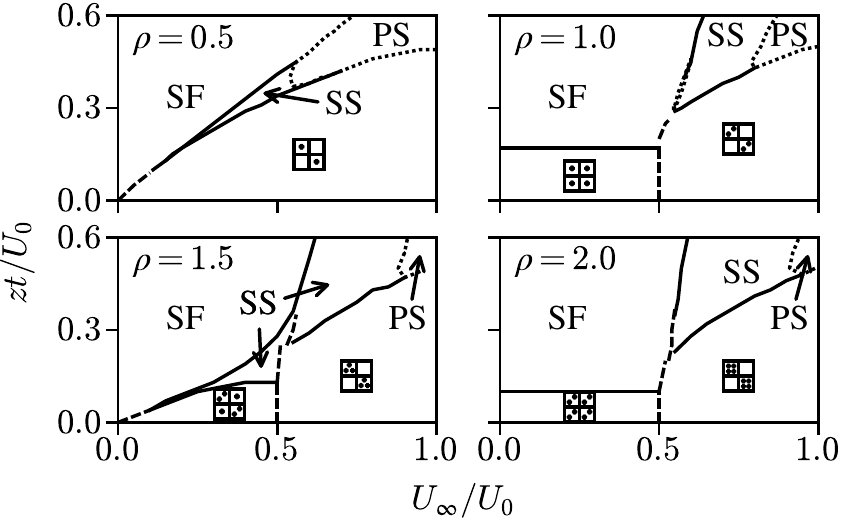}
  \caption{
  Ground state phase diagram as in Fig.~\ref{fig:cavity-ground-state-phase-diagram} in the $U_\infty/U_0-t/U_0$ plane and for fixed density. The subplots correspond to (from left to right) upper panel: $\rho=1/2$, $\rho=1$, lower panel: $\rho=3/2$, $\rho=2$. 
  See text for details.
  % See files cavity-meanfield-groundstate-independent-density-2018-09-24T12:55:01+0200.json.xz cavity-meanfield-groundstate-independent-density-2018-09-25T07:05:07+0200.json.xz
  Dashed (solid) lines at the phase boundaries indicate discontinuous (continuous) variation of the order parameters  Dotted lines show the boundary of the PS regions.
  In the numerical implementation, the cutoff to the site occupation is set at $n_{\text{max}} = 31$ and we took 175 initial guesses.
  The accuracy in the determination of the mean-field order parameters is $\varepsilon = 10^{-6}$, the precision in the determination of the density is $\varepsilon_{\rho} = 10^{-4}$. Details regarding the phase diagram at densities $\rho = 0.5$ and $\rho = 1$ are given in Appendix \ref{app:constant-density}.  }
  \label{fig:cavity-ground-state-constant-density}
\end{figure}

Figure~\ref{fig:cavity-ground-state-constant-density} shows the phase diagram for $\rho=1/2,1,3/2$, and $2$. For $\rho = 1/2$ there is no MI phase. Nevertheless, for $U_\infty>0$, we observe parameter regions where the ground state is in the CDW phase, corresponding to the occupation $\{ 0, 1 \}$ between neighbouring sites. For $U_\infty \lesssim 0.1 U_0$  CDW and SF are separated by a first order phase transition. This phase boundary is characterized by a discontinuity of the population imbalance, the transition line is at a value of the tunneling rate which scales seemingly linearly with $U_\infty/U_0$ and ends at a tricritical point. After this point the SS phase separates the CDW from the SF phase and the order parameters vary continuously at the transition lines separating SF-SS and SS-CDW. The area enclosed by the dotted lines in the diagram is the parameter region where we could not find any data point, namely, where there is no value of the chemical potential corresponding to $\rho=1/2$. We denote this region by Phase Separation (PS), after observing that simulations for these values  in a canonical ensemble using QMC reported  negative compressibility and have been linked to a phase separation between the SF and SS phases \cite{flottat-et-al-2017}.

We first notice that this phase diagram coincides with the one reported in Ref.~\cite{dogra-et-al-2016}, apart from the fact that the authors seem to always find a SS phase separating the CDW and the SF phases, and thus they report neither a direct SF-CDW transition nor a PS region. In particular, all transitions they find for $\rho=1/2$ are of second order. This difference, and especially the absence of the PS region, might be attributed to  different methods for determining the ground state at a fixed density in a grand-canonical ensemble. The authors of Ref.~\cite{dogra-et-al-2016} first identify the states at the target density for given $t,U_\infty$, and then search for the lowest-energy state in this set \cite{dogra-et-al-2016, dogra-2017:private-communication}.
In our work, instead, we first determine the states at the lowest energy as a function of $\mu$ for given $t,U_\infty$. In this set of states we then search for the one corresponding to the target density.
The PS region corresponds to the parameters for which the target density cannot be found. 
Details of our analysis are reported in Appendix \ref{app:constant-density}.

Remarkably, the plot for $\rho=1/2$ reproduces qualitatively the  corresponding diagram obtained with QMC in Ref. \cite{flottat-et-al-2017}. In particular, the authors claim to find a direct transition CDW-SF at smaller values of $t/U_0$ (larger values of $U_0/t$), however they cannot determine its nature due to the fact that the QMC simulations are not conclusive in this parameter regime. The salient difference with our result is  that the authors do not report stable SS phases for $\rho=1/2$. This does not exclude, in our view, that a stable SS phase could exist in a small parameter region close to the tricritical point, which might have not been included in the data sampling.

The phase diagrams for $\rho = 1$ and $\rho=2$ have a similar structure. For both cases the phases are MI, SF, SS, and CDW with maximal population imbalance.
The MI-SF and the SS-CDW transitions are continuous. The MI-CDW, instead, is a discontinuous transition. Moreover, for both densities $\rho=1,2$ there is a direct, discontinuous transition CDW-SF at $U_\infty\sim 0.5 U_0$ which ends at a tricritical point. As $U_\infty/U_0$ is further increased, this transition line splits into two phase boundaries: the SS-CDW and the SF-SS.
%Here, it splits into two phase boundaries: the SS-CDW and the SF-SS.
The SF-SS transition is continuous except for a small region close to the tricritical point. This region corresponds to the parameter regime for which we find no solution of the equation $\rho(\mu/U_0,\tilde t,\tilde U_\infty)=1$. In the case of $\rho=2$, instead, the   
 SF-SS transition is discontinuous close to the tricritical point. On the other hand, the SS-CDW is a continuous transition until a critical value $U_\infty(t,\rho)$, after which we find a PS region.

%A phase diagram for $\rho=1$ was also reported in Ref.~\cite{dogra-et-al-2016}. Our diagram in Fig. \ref{fig:cavity-ground-state-constant-density} and their diagram are in full agreement in the parameter intervals considered in Ref.~\cite{dogra-et-al-2016}. 
The diagram for $\rho=1$ in Fig. \ref{fig:cavity-ground-state-constant-density} is in full agreement with the one reported in Ref.~\cite{dogra-et-al-2016}, within the parameter intervals considered.
Moreover, it also agrees qualitatively with the phase diagram evaluated using QMC  \cite{flottat-et-al-2017}, apart for two salient features: The authors do not report a PS and the transition line SS-SF is continuous along the whole branch of their phase diagram.  We note that the phase diagram at $\rho=1$  in Fig. \ref{fig:cavity-ground-state-constant-density} is similar to the one of Ref.~\cite{wald-et-al-2018}, obtained by minimizing the mean-field free energy of a canonical ensemble in a constrained Hilbert space. According to the free energy landscape of Ref.~\cite{wald-et-al-2018}, the MI-CDW transition is characterized by a large parameter region where the two phases are metastable.

The phase diagram at $\rho=3/2$ in Fig. \ref{fig:cavity-ground-state-constant-density} exhibits a CDW phase with $\{ 1, 2 \}$, separated by a discontinuous transition to the CDW phase with $\{ 3, 0 \}$ at $U_\infty=U_0/2$.
The CDW $\{ 1, 2 \}$ emerges at infinitesimally small tunneling parameters, it has a first-order transition to a SF until a finite value
$U_\infty< U_0/2. $ %$U_\infty<U_0/2$. 
This CDW$\{1,2\}$-SF phase boundary ends at a tricritical point, after which SF and CDW are separated by a SS phase. The transition SF-SS is continuous in the whole parameter range. On the other hand, SS-CDW$\{3,0\}$ transition, becomes discontinuous for a small interval of values about $U_\infty\sim 0.5 U_0$.
Within the SS phase, moreover, there is a discontinuous transition at  $U_\infty=U_0/2$ where the population imbalance undergoes a jump from $|\theta| \approx 1$ to $|\theta| \approx 3$.
This jump was reported also in Ref. \cite{dogra-et-al-2016}. Instead, QMC studies found it to be a crossover  \cite{flottat-et-al-2017}.
Moreover the direct transition between SF and CDW $\{1,2\}$ seems to not have been found by static mean field calculations in Ref. \cite{dogra-et-al-2016}. QMC simulations \cite{flottat-et-al-2017} here reported this direct transition, however they could not determine its order.
Finally, we notice a region for strong long-range interaction and large tunneling where no solution exists and which was not reported by static mean-field calculations \cite{dogra-et-al-2016}.

We note that the phase diagram for $\rho =1/2$ is similar to that of the extended Bose-Hubbard model with repulsive nearest neighbour interaction: For nearest-neighbour interactions and small tunneling rates Quantum Monte Carlo simulations report no stable SS phase but a direct transition CDW-SF \cite{batrouni-et-al-1995, batrouni-scalettar-2000, sengupta-et-al-2005}, which is found to be of first order \cite{batrouni-et-al-1995, batrouni-scalettar-2000}. Furthermore in the intermediate tunneling regime a SF-SS and SS-CDW transition is observed \cite{kimura-2011, ohgoe-et-al-2012}.
For $\rho = 1$, in the extended Bose-Hubbard model with repulsive nearest neighbour interaction and small tunneling rates there is a MI-CDW transition \cite{sengupta-et-al-2005, kimura-2011}, while the author of ref.~\cite{kimura-2011} finds a SF-CDW transition for an intermediate tunneling rate $zt = 0.3 U_0$, and SF-SS and SS-CDW transitions for a very large tunneling rate $z t = U_0$.
However, no PS at $\rho = 1$ and $\rho = 1/2$ is reported in Refs. \cite{kimura-2011, ohgoe-et-al-2012}.

\section{Conclusions}
\label{Sec:5}

We have performed a mean-field analysis of the phase diagram of the extended Bose-Hubbard model, where the bosons have a repulsive contact interaction and experience an infinitely long-range two-body potential. The systematic comparison between the phase diagram obtained for the cavity Bose-Hubbard model and the one for repulsively interacting dipolar gases in two dimensions shows clear differences already within the mean-field treatment, such as for instance the direct first-order transition CDW-MI at a critical value of the chemical potential, that is absent for the dipolar case. The ground-state phase diagram we calculate mostly agrees with the static mean-field diagram of Ref. \cite{dogra-et-al-2016}. There are two important differences: differing from Ref. \cite{dogra-et-al-2016}, for the density $\rho=1/2$ and $\rho=3/2$ we predict a direct transition between Superfluid (SF) and Charge-Density Wave (CDW), which is first order for the densities we considered. Moreover, in the region where the authors of  Ref. \cite{dogra-et-al-2016} predict stable lattice supersolid (SS) phases, we find also regions where instead there is a Phase Separation (PS).
We attribute these discrepancies to different methods for determining the ground state at fixed density from the grand-canonical ensemble calculations, as we detailed in Sec. \ref{Sec:Density}.

The stability of the SS phase has been extensively analysed by means of Quantum Monte Carlo methods for a canonical and a grand-canonical ensemble in Ref. \cite{flottat-et-al-2017}. Our diagrams and the diagrams of Ref. \cite{dogra-et-al-2016} at fixed densities, extracted from grand-canonical ensemble calculations, are in remarkable qualitative agreement with the QMC diagrams in the interval of parameters where the QMC diagram have been determined. The discrepancies regard the stability of the SS regions at fixed densities. These discrepancies could be due to the fact that the QMC collected data did not sample the regions where these differences are found (as one could conjecture by taking the parameters reported in Ref. \cite{flottat-et-al-2017} for which the stability of the SS phase was analysed and mapping them into our phase diagram). Most probably, the discrepancy arises from the fact that our static mean-field approach cannot appropriately take into account the interplay between strong long-range interactions and quantum fluctuations. %This interplay is in fact largely unexplored and its role at criticality still needs to be characterized.

\acknowledgments
The authors are thankful to George Batrouni, Nishant Dogra, Rosario Fazio, Gabriel Landi, Astrid Niederle, Heiko Rieger, Katharina Rojan, and Sascha Wald, for discussions.
Financial support by the DFG priority program no.~1929 GiRyd, by the DFG DACH program "Quantum crystals of matter and light", and by the German Ministry of Education and Research (BMBF) via the Quantera project ``NAQUAS''.
Project NAQUAS has received funding from the QuantERA ERA-NET Cofund in Quantum Technologies implemented within the European Union's Horizon 2020 Programme.
CC acknowledges funding from the Alexander-von-Humboldt Foundation for her visit to Saarland University and grant BID-PICT 2015-2236.

\appendix

\section{Numerical calculation of the ground state}
\label{app:numerical-ground-state}

In this Appendix, we describe the algorithm used to find the self-consistent ground state of the local mean field Hamiltonians \eqref{eq:mean-field-hamiltonians}.
We measure all physical parameters of the Hamiltonian in units of the on-site interaction, $\tilde{\mu} = \mu / U_0$, $\tilde{U}_\infty = U_\infty / U_0$, $\tilde{t} = z t / U_0$ and obtain the Hamiltonians
\begin{alignat}{2}
  \tilde{H}_{\text{e}} & = - \tilde{t} \varphi_{\text{o}} ( a + a^\dagger - \varphi_{\text{e}} ) + \frac{1}{2} n (n - 1) - \tilde{U}_\infty \theta n + \frac{\tilde{U}_\infty}{4} \theta^2 - \tilde{\mu} n \text{,} \\
  \tilde{H}_{\text{o}} & = - \tilde{t} \varphi_{\text{e}} ( a + a^\dagger - \varphi_{\text{o}} ) + \frac{1}{2} n (n - 1) + \tilde{U}_\infty \theta n + \frac{\tilde{U}_\infty}{4} \theta^2 - \tilde{\mu} n \text{,}
  \label{eq:hamiltonians-numerical}
\end{alignat}
with the same eigenenergies and -states as the Hamiltonians \eqref{eq:mean-field-hamiltonians}.
We fix the parameters of the Hamiltonian $\tilde{t}$, $\tilde{U}_\infty$ and $\tilde{\mu}$. The mean-field order parameters $\varphi_{\text{e}}$, $\varphi_{\text{o}}$ and $\theta$ are now the free variables. The problem is formulated as follows. We first introduce the function
\begin{equation}
  f(\varphi_{\text{e}}, \varphi_{\text{o}}, \theta) = ( \langle a \rangle_{\text{e}}, \langle a \rangle_{\text{o}}, \langle n \rangle_{\text{e}} - \langle n \rangle_{\text{o}} ) \text{,}
  \label{eq:mean-field-iteration-f}
\end{equation}
where $\langle \cdot \rangle_{s}$ denotes the single-site expectation value with respect to the ground state of the Hamiltonian $H_s(\varphi_{\text{e}}, \varphi_{\text{o}}, \theta)$, for $s \in \{ \text{e}, \text{o} \}$.
Further, we define $F$ to be the set of fixed points of $f$,
\begin{equation}
  F = \left\{ (\varphi_{\text{e}}, \varphi_{\text{o}}, \theta) : f(\varphi_{\text{e}}, \varphi_{\text{o}}, \theta) = (\varphi_{\text{e}}, \varphi_{\text{o}}, \theta) \right\} \text{.}
\end{equation}
The goal is to find the self-consistent order parameters which minimize the energy per site,
\begin{equation}
  (\varphi_{\text{e}}, \varphi_{\text{o}}, \theta) = \operatorname*{argmin}\limits_{(\varphi_{\text{e}}, \varphi_{\text{o}}, \theta) \in F} \left\{ \frac{1}{2} \left( \langle H_{\text{e}} \rangle_{\text{e}} + \langle H_{\text{o}} \rangle_{\text{o}} \right) \right\} \text{.}
\end{equation}

The basic idea of the algorithm is that of fixed point iteration:
Apply $f$ repeatedly to some random $(\varphi_{\text{e}}, \varphi_{\text{o}}, \theta)$, until applying it again does not significantly change the input \cite{habibian-et-al-2013, dhar-et-al-2011, niederle-morigi-rieger-2016}.

We measure the distance between mean-field order parameters by the infinity-norm and relax the criterion for $(\varphi_{\text{e}}, \varphi_{\text{o}}, \theta)$ to be a fixed-point to
{\footnotesize{
\begin{equation}
  \left\lVert (\varphi_{\text{e}}, \varphi_{\text{o}}, \theta) - (\varphi_{\text{e}}', \varphi_{\text{o}}', \theta') \right\rVert_\infty = \operatorname{max}(|\varphi_{\text{e}} - \varphi_{\text{e}}'|, |\varphi_{\text{o}} - \varphi_{\text{o}}'|, |\theta - \theta'|) < \varepsilon \text{,}
\end{equation}}}
where $(\varphi_{\text{e}}', \varphi_{\text{o}}', \theta') = f(\varphi_{\text{e}}, \varphi_{\text{o}}, \theta)$, and $\varepsilon$ is some predefined tolerance, e.~g.\ $\varepsilon = 10^{-6}$.

This na\"{i}ve algorithm has the following problems, however:
First, if the algorithm converges to some point, there is no guarantee that this point minimizes the energy per site.
Second, the algorithm is not guaranteed to converge.
Third, the algorithm sometimes converges sublinearly, and thus extremely slowly.

We approach the first problem by taking a sufficient large number of initial guesses.% and finally taking only the result which leads to the lowest energy.
 We always deterministically take the following $75$ initial guesses: $(\varphi_{\text{e}}, \varphi_{\text{o}}, \theta) \in \cup_{\{ n \in 0, \ldots, 24 \} } \{ (0, 0, n), (0.001, 0.002, n), (0.1, 0.2, n) \}$.
Additionally, we use the Mersenne Twister and Ranlux48 algorithms to pseudorandomly sample initial guesses from the $\mathrm{Cauchy}(0, 1)$ distribution.
We find that $50$ random initial values are sufficient to find the minimal energy and verify this by taking more initial guesses and verifying that the energy does not decrease significantly.

The problem that the algorithm sometimes does not converge manifests in the way of cycles of the form that $f(\varphi_{\text{e}}, \varphi_{\text{o}}, \theta) \approx (-\varphi_{\text{e}}, -\varphi_{\text{o}}, -\theta)$ and $f(-\varphi_{\text{e}}, -\varphi_{\text{o}}, -\theta) \approx (\varphi_{\text{e}}, \varphi_{\text{o}}, \theta)$.
We detect this by comparing not only the mean-field order parameters, but also their absolute values.
If the difference of the absolute values is smaller than $\varepsilon / 10$ for $1000$ consecutive iterations, we re-run the algorithm with the absolute values of the final order parameters $(\varphi_{\text{e}}', \varphi_{\text{o}}', \theta')$ as an initial guess.
To ensure that we still find the minimal energy, we compare the energies of the result of the initial run $(\varphi_{\text{e}}', \varphi_{\text{o}}', \theta')$, with that of the second run, $(\varphi_{\text{e}}'', \varphi_{\text{o}}'', \theta'')$.
For this comparison, we do not consider the eigenvalues of the two Hamiltonians, but the expectation value of the updated Hamiltonian with respect to the ground state of the Hamiltonian before the update.
If the energy of the second run is smaller, we accept this solution, otherwise we reject it.

\begin{figure}
  \includegraphics{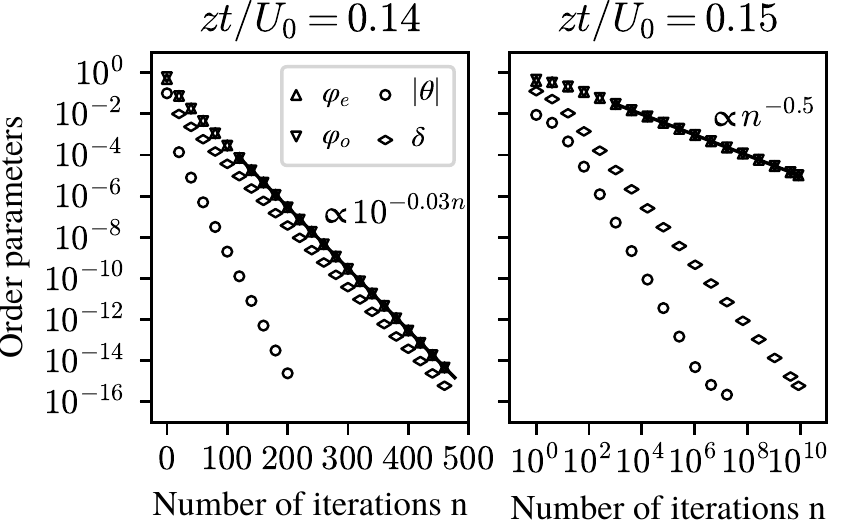}
  \caption{
  Convergence of the numerical mean-field iterative algorithm as a function of the number of iterations (or applications of $f$, as given in equation \eqref{eq:mean-field-iteration-f}.
  In both cases, the occupation is cut off above $n_{\text{max}} = 31$, $\tilde{U_\infty} = 0.26$, $\tilde{\mu} = 0.6$, and the initial guess is $(\varphi_{\text{e}}, \varphi_{\text{o}}, \theta) = (0.5, 0.6, 0.1)$.
  On the left, $\tilde{t} = 0.14$ and the convergence is linear.
  On the right, $\tilde{t} = 0.15$ and the convergence is sublinear.
  Both points are close to the MI-SF phase boundary.
  The markers show the order parameters and the maximum difference $\delta$ of the order parameters from one iteration to the next.
  The black lines are obtained by linear regression of an exponential function (left) and a power function (right) to the maximum order parameter.
  The calculations took around $6\,\text{CPU-days}$ on a Intel Core i7-2600 CPU at a clock rate of $\sim 3.6\,\text{GHz}$.
  }
  \label{fig:numerical-convergence}
\end{figure}

Finally, we note that the algorithm converges to the set tolerance within a few hundred or thousand iterations (i.~e.\ applications of $f$) in a large region of the phase diagram.
In this case, the algorithm converges linearly.
However, in some cases it converges sublinearly and extremely slowly.
Figure \ref{fig:numerical-convergence} shows a comparison of two cases, for two points in the phase diagram which are close to each other, and identical initial guesses.

The algorithm converges that slow only at relatively few points in the phase diagram.
We have verified that the number of iterations does not strongly influence the value of the resulting mean-field order parameters, by comparing the results after $10^4$ and $10^6$ iterations.
%We conclude that we reliably find the mean-field ground state.

For finding the ground state of the Hamiltonians \eqref{eq:hamiltonians-numerical}, we truncate the Hilbert space of each site taking the cutoff $n_{\text{max}} = 31$, leaving us with two tridiagonal real symmetric $32 \times 32$ matrices (in the Fock basis), which we diagonalize numerically.
We identify the cutoff $n_{\text{max}} = 31$ by performing calculations also for $n_{\text{max}} = 23$ and $n_{\text{max}} = 63$ and verifying that the results do not differ significantly.

We implemented the algorithm in the C++ language, using the GNU compiler collection (versions 7.2 and 8) and clang (versions 5 and 6).
For diagonalization, we used the library Eigen3.
We verified the self-consistency of the results with a partial implementation of the algorithm in the Python language (version 3.6) and using the numpy library (version 1.14.5) for diagonalization.
\cite{software}

\section{Supersolidity and phase separation for fixed densities}
\label{app:constant-density}

In this Appendix, we report details of the calculations for determining the phase diagram for constant densities of Fig.~\ref{fig:cavity-ground-state-constant-density}.

We obtain the order parameters for fixed densities by adjusting the chemical potential such that the ground state has the target density.
More precisely, we perform a bisection algorithm starting at $\mu = - U_0$ which gives a density lower than the target density and $\mu = 3 U_0$ which gives a density higher than the target density.
In every step, the ground state for the midpoint of the $\mu$ interval is calculated following the procedure detailed in Appendix \ref{app:numerical-ground-state}, its density is computed, and the interval is halved such that at the lower (upper) point of the interval, the density is smaller (larger) than the target density.
We repeat this up to twenty times until either the target density is reached or we conclude that a solution is not possible.
When the density is not attained with the required precision, which we set to $\varepsilon_\rho = 10^{-4}$, we name the corresponding point of the phase diagram phase separation, following Ref.~\cite{flottat-et-al-2017}.
Otherwise, we determine the phase from the values of the order parameters $\theta$ and $\varphi$.

\begin{figure}
  \includegraphics{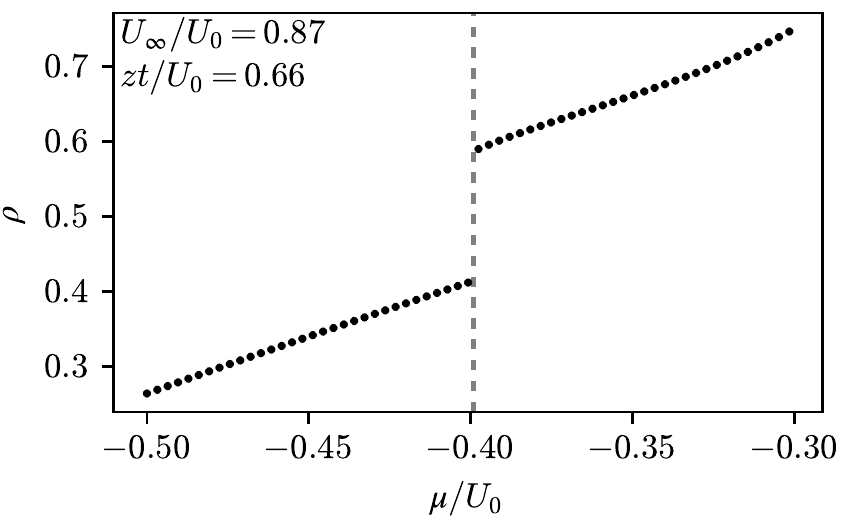}
  % file cavity-meanfield-groundstate-independent-hamiltonian-parameters-2018-12-04T16:20:07+0100.json.xz
  \caption{
  Density $\rho$ as a function of the chemical potential $\mu$ for $U_\infty / U_0 = 0.87, z t / U_0 = 0.66$.
  The vertical bar marks a jump in the $\rho(\mu)$ curve.
  }
  \label{fig:ps-at-density-0.5}
\end{figure}

\begin{figure}
  \includegraphics{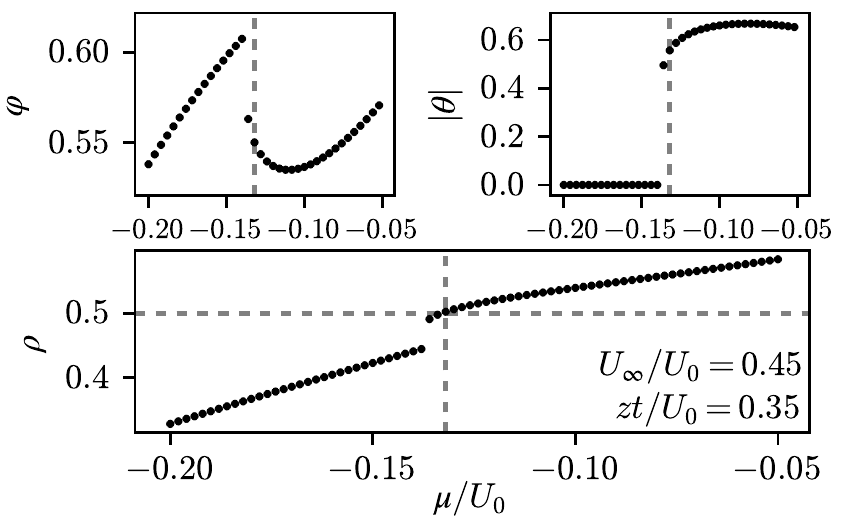}
  % file cavity-meanfield-groundstate-independent-hamiltonian-parameters-2018-06-21T10:49:25+0200.json.xz
  \caption{
  (c) Density $\rho$ as a function of the chemical potential $\mu$ for $U_\infty / U_0 = 0.45, z t / U_0 = 0.35$.
  The horizontal bar marks the density $\rho = 0.5$.
  (a) Superfluid order paramater $\varphi(\mu)$ for the same parameters.
  (b) Even-odd imbalance $|\theta(\mu)|$ for the same parameters.
  The vertical bar marks $\rho(\mu) = 0.5$.
  }
  \label{fig:ss-at-density-0.5}
\end{figure}

For a density of $\rho = 0.5$, we find both a SS region and a region of phase separation, unlike Refs.~\cite{dogra-et-al-2016, flottat-et-al-2017}.
Fig.~\ref{fig:ps-at-density-0.5} shows the $\rho(\mu)$ curve for a PS point.
Fig.~\ref{fig:ss-at-density-0.5} shows the $\rho(\mu)$ curve for a different point, where the density $\rho(\mu \approx -0.132) = 0.5$.
The same figure shows the superfluid order parameter and the even-odd imbalance; for the parameters where $\rho(\mu) = 0.5$, the ground state is SS.

\begin{figure}
  \includegraphics{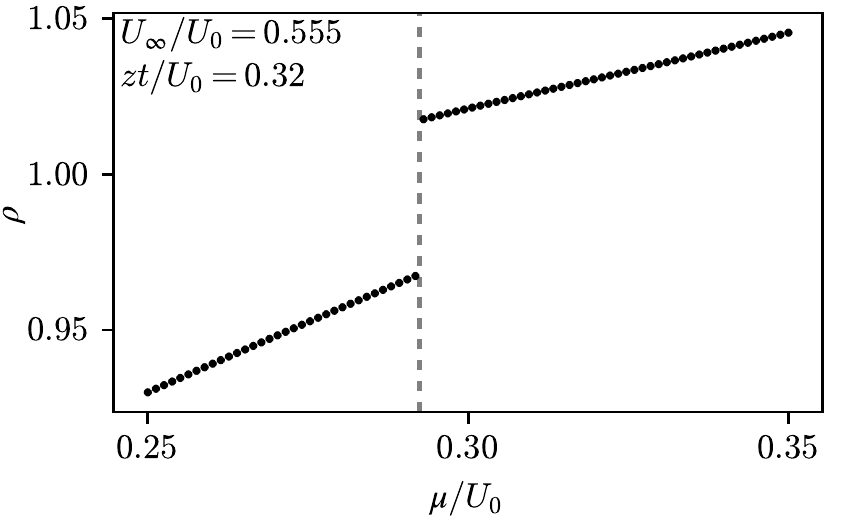}
  % cavity-meanfield-groundstate-independent-hamiltonian-parameters-2018-12-14T14:04:56+0100.json.xz
  \caption{
  Density $\rho$ as a function of the chemical potential $\mu$ for $U_\infty / U_0 = 0.555$, $z t / U_0 = 0.320$.
  The vertical bar marks a jump in the $\rho(\mu)$ curve.
  }
  \label{fig:ps-at-density-1.0}
\end{figure}

In Fig.~\ref{fig:ps-at-density-1.0}, we show the $\rho(\mu)$ curve for a PS point for density $\rho = 1$, which was not reported so far \cite{dogra-et-al-2016, flottat-et-al-2017}.
As shown in fig.~\ref{fig:cavity-ground-state-constant-density}, we also find a PS region for $\rho = 1.5$, which was not reported so far \cite{dogra-et-al-2016, flottat-et-al-2017}.

\begin{figure}
  \includegraphics{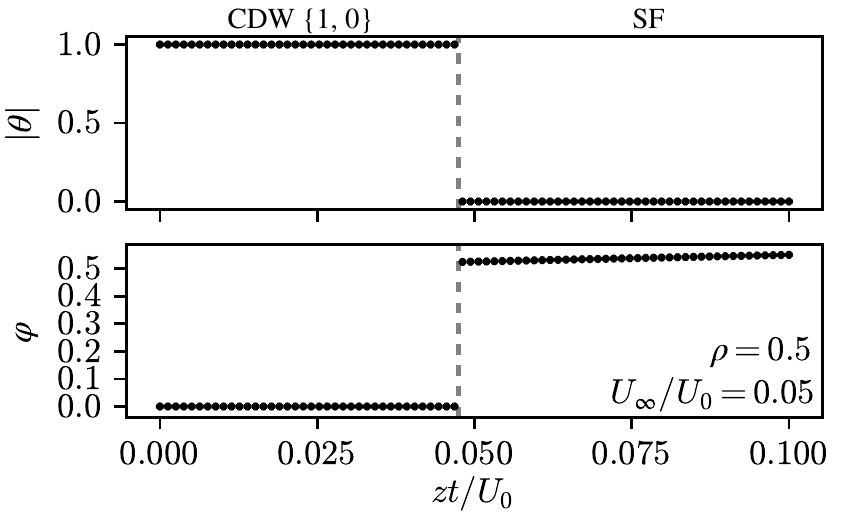} \\
  % cavity-meanfield-groundstate-independent-density-2018-12-12T17:09:16+0100.json.xz
  \includegraphics{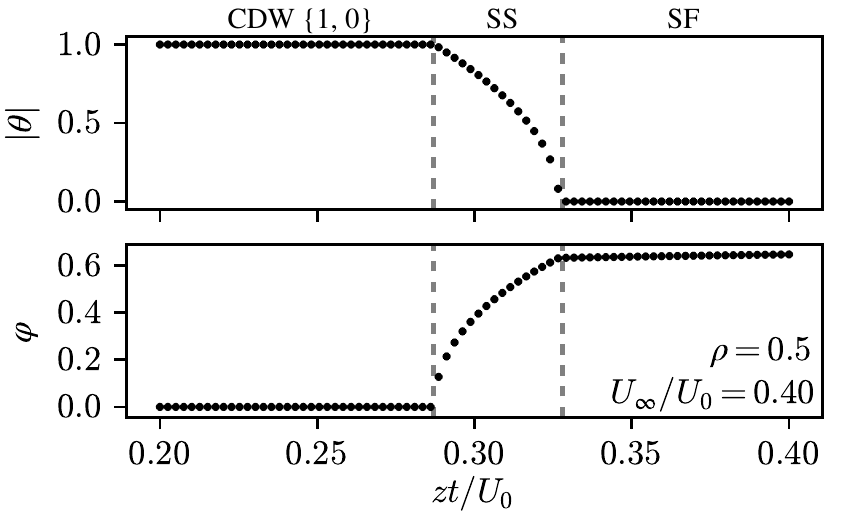} \\
  % cavity-meanfield-groundstate-independent-density-2018-12-13T17:15:07+0100.json.xz
  \includegraphics{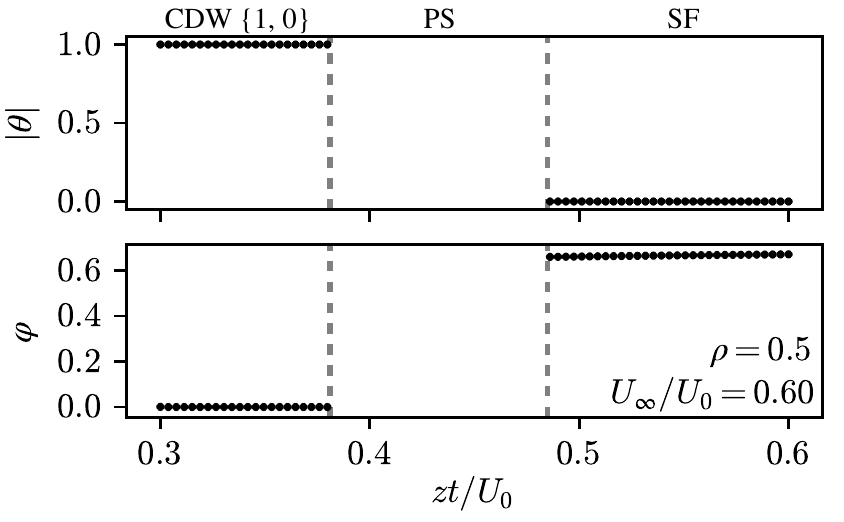}
  % cavity-meanfield-groundstate-independent-density-2018-12-13T18:11:11+0100.json.xz
  \caption{
  Imbalance $|\theta|$ and superfluid order parameter $\varphi$ for cuts of the phase diagram \ref{fig:cavity-ground-state-constant-density} with a constant density of $\rho = 0.5$.
  The vertical dashed lines show the phase transition points.
  The phases are indicated by the labels above the plots.
  }
  \label{fig:density-0.5-cuts}
\end{figure}

\begin{figure*}
  \includegraphics{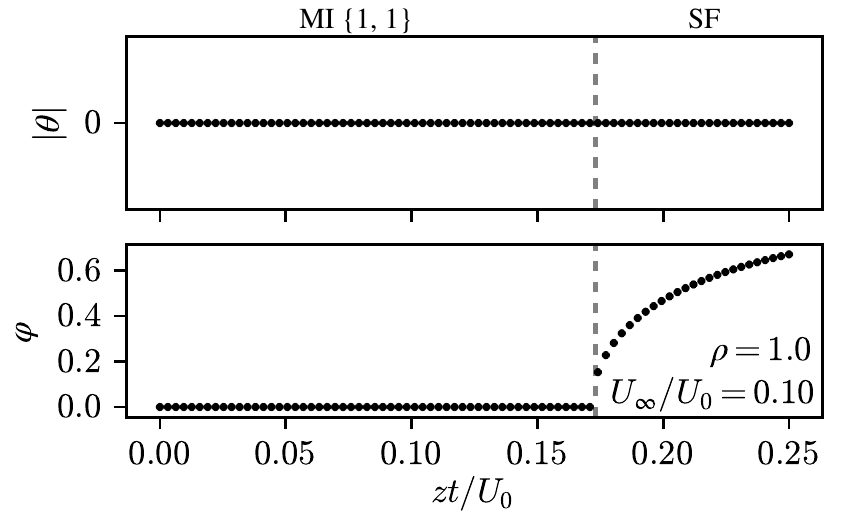}
  % cavity-meanfield-groundstate-independent-density-2018-12-14T08:22:31+0100.json.xz
  \includegraphics{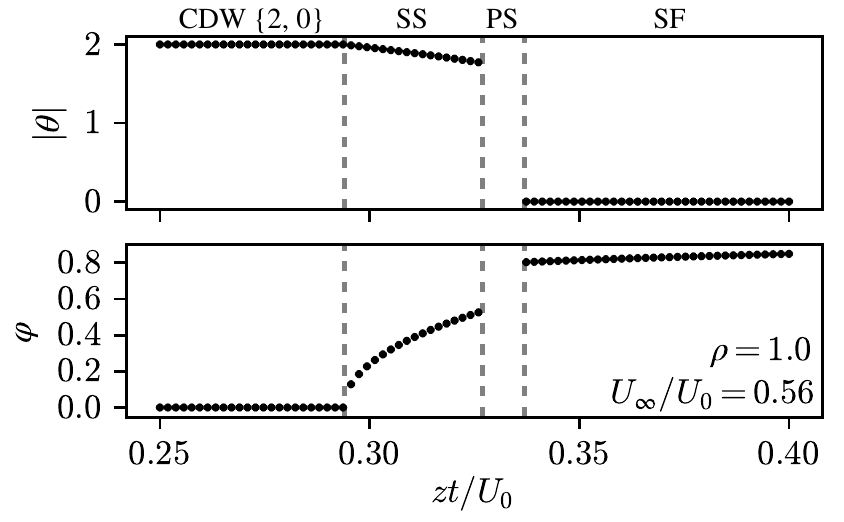} \\
  % cavity-meanfield-groundstate-independent-density-2018-12-14T09:12:24+0100.json.xz
  \includegraphics{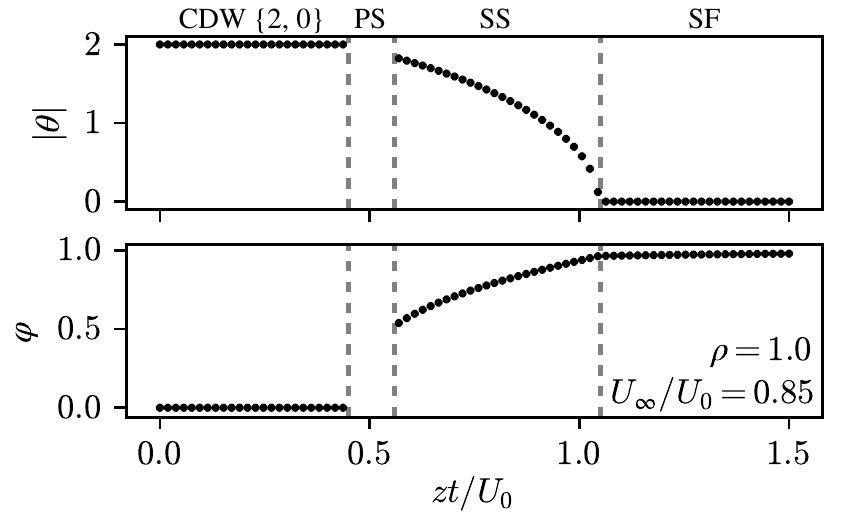}
  % cavity-meanfield-groundstate-independent-density-2018-12-14T10:46:08+0100.json.xz
  \includegraphics{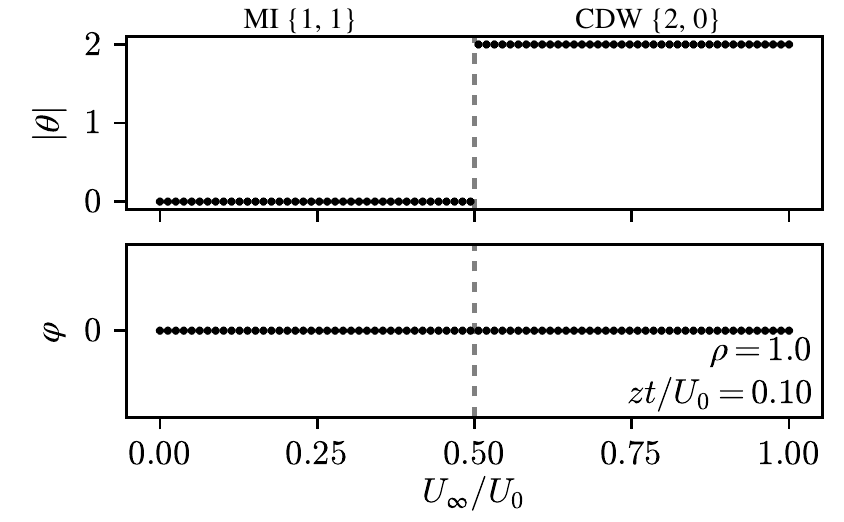} \\
  % cavity-meanfield-groundstate-independent-density-2018-12-14T11:45:50+0100.json.xz
  \includegraphics{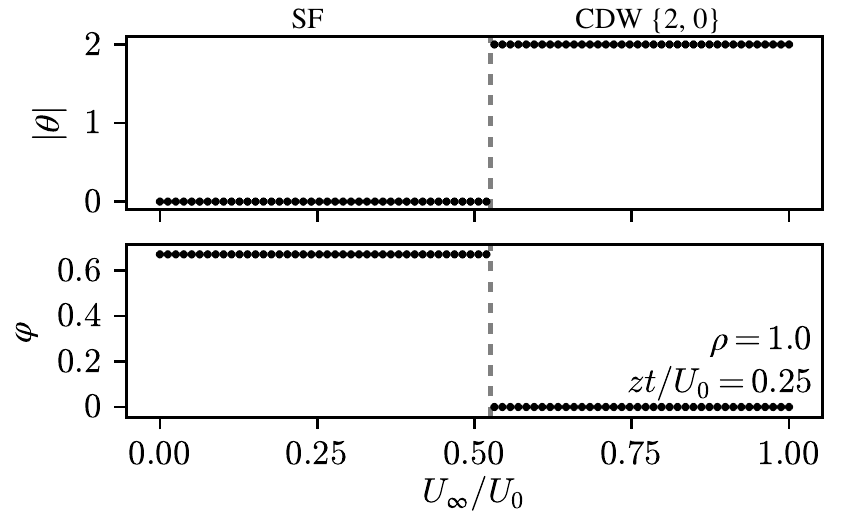}
  \caption{
  Imbalance $|\theta|$ and superfluid order parameter $\varphi$ for cuts of the phase diagram \ref{fig:cavity-ground-state-constant-density} with a constant density of $\rho = 1$.
  The vertical dashed lines show the phase transition points.
  The phases are indicated by the labels above the plots.
  }
  \label{fig:density-1-cuts}
\end{figure*}

In Fig.~\ref{fig:density-0.5-cuts}, we show the order parameters along cuts of the phase diagram in figure \ref{fig:cavity-ground-state-constant-density}, specifically the density $\rho = 0.5$.
We show similar plots for the density $\rho = 1$ in figure \ref{fig:density-1-cuts}.
We used plots similar to the ones shown in Figs.~\ref{fig:density-0.5-cuts} and \ref{fig:density-1-cuts} to determine all phase boundaries of Fig.~\ref{fig:cavity-ground-state-constant-density}.

\end{document}